\documentclass[authorversion,sigconf]{acmart}

\usepackage[skip=4pt]{caption}
\usepackage[caption=false]{subfig}
\usepackage{bytefield}
\usepackage{balance}
\usepackage{xcolor}
\usepackage{xspace}
\usepackage{url}
\usepackage{hyperref}
\usepackage{graphicx}
\usepackage[binary-units=true]{siunitx}

\newcommand{\code}[1]{\texttt{#1}}

\newcommand{\ie}{i.e.,\xspace}
\newcommand{\eg}{e.g.,\xspace}
\newcommand{\Eg}{E.g.,\xspace}
\newcommand{\cf}{cf.\xspace}
\newcommand{\etal}{et al.\xspace}

\usepackage{enumitem}
\setlist[itemize]{label=\textbullet,topsep=0pt,leftmargin=1em}

\DeclareSIUnit\usd{USD}
\DeclareSIUnit\bitcoin{BTC}
\DeclareSIUnit\satoshi{satoshi}
\DeclareSIUnit\btcweight{WU}
\DeclareSIUnit{\million}{\text{million}}

\newcommand{\rulesep}{\unskip\ \vrule\ }

\newcommand\shorturl[1]{\href{http://#1}{\nolinkurl{#1}}}
\newcommand\shorturls[1]{\href{https://#1}{\nolinkurl{#1}}}

\newcommand{\name}{AnonBoot\xspace}

\newcounter{goal}

\newcommand{\gref}[1]{{\normalfont \textbf{(\ref{#1})}}}
\newcommand{\step}[2][]{\raisebox{.5pt}{Step~\textcircled{\raisebox{-.5pt}{\small#2\raisebox{0.5pt}{\footnotesize #1}}}}}

\copyrightyear{2020}
\acmYear{2020}
\setcopyright{acmcopyright}
\acmConference[ASIA CCS '20]{Proceedings of the 15th ACM Asia Conference on Computer and Communications Security}{October 5--9, 2020}{Taipei, Taiwan}
\acmBooktitle{Proceedings of the 15th ACM Asia Conference on Computer and Communications Security (ASIA CCS '20), October 5--9, 2020, Taipei, Taiwan}
\acmPrice{15.00}
\acmDOI{10.1145/3320269.3384729}
\acmISBN{978-1-4503-6750-9/20/10}

\begin{document}

\fancyhead{}
\title{Utilizing Public Blockchains for the Sybil-Resistant Bootstrapping of Distributed Anonymity Services}

\author{Roman Matzutt, Jan Pennekamp, Erik Buchholz, Klaus Wehrle}
\email{{matzutt, pennekamp, buchholz, wehrle}@comsys.rwth-aachen.de}
\affiliation{%
    \institution{Communication and Distributed Systems, RWTH Aachen University, Germany}
}
\renewcommand{\shortauthors}{Matzutt et al.}

\begin{abstract}
Distributed anonymity services, such as onion routing networks or cryptocurrency tumblers, promise privacy protection without trusted third parties.
While the security of these services is often well-researched, security implications of their required bootstrapping processes are usually neglected:
Users either jointly conduct the anonymization themselves, or they need to rely on a set of non-colluding privacy peers.
However, the typically small number of privacy peers enable single adversaries to mimic distributed services.
We thus present \emph{\name}, a Sybil-resistant medium to securely bootstrap distributed anonymity services via public blockchains.
\name enforces that peers periodically create a small proof of work to refresh their eligibility for providing secure anonymity services.
A pseudo-random, locally replicable bootstrapping process using on-chain entropy then prevents biasing the election of eligible peers.
Our evaluation using Bitcoin as \name's underlying blockchain shows its feasibility to maintain a trustworthy repository of \num{1000} peers with only a small storage footprint while supporting arbitrarily large user bases on top of most blockchains.
\end{abstract}

\ccsdesc[500]{Security and privacy~Pseudonymity, anonymity and untraceability}
\ccsdesc[500]{Networks~Peer-to-peer protocols}

\keywords{anonymization; bootstrapping; public blockchain; Sybil attack; anonymity network; cryptocurrency tumbler; Bitcoin; Tor}

\maketitle

\section{Introduction}
\label{sec:introduction}

Preserving user privacy on the Internet has become a complex task due to increasingly pervasive measures for online surveillance:
While re-establishing their anonymity traditionally was only crucial for a set of especially privacy-aware users, the Snowden revelations have shown that every online user's privacy is at stake~\cite{2013_landau_snowden}.
This shift further fueled distributed anonymity services, such as message shuffling networks~\cite{1981_chaum_mixnet}, anonymous communication networks based on onion routing~\cite{2004_dingledine_tor}, or cryptocurrency tumblers~\cite{2015_ziegeldorf_coinparty,2018_ziegeldorf_coinpartyv2,2018_meiklejohn_moebius}.
While various works have investigated secure building blocks for anonymity services, those works typically overlook the bootstrapping of such services.
Often, related work simply assumes non-colluding peers, \eg because of their operators' presumed real-world reputation.
However, this perceived reputation does not always warrant trust, as evidenced, \eg by numerous alleged scams regarding cryptocurrencies~\cite{2018_ziegeldorf_coinpartyv2,2014_badbitcoin_scams,2015_vasek_scams} and the need for manually reporting~\cite{2014_tor_report} or actively probing~\cite{2011_chakravarty_tor_snooping,2014_winter_spoiled_tor} bad peers in the Tor network.
Hence, the question remains:
\emph{How to securely bootstrap distributed anonymity services without having to rely on operator reputation?}

\begin{figure}[t]
    \centering
    \includegraphics{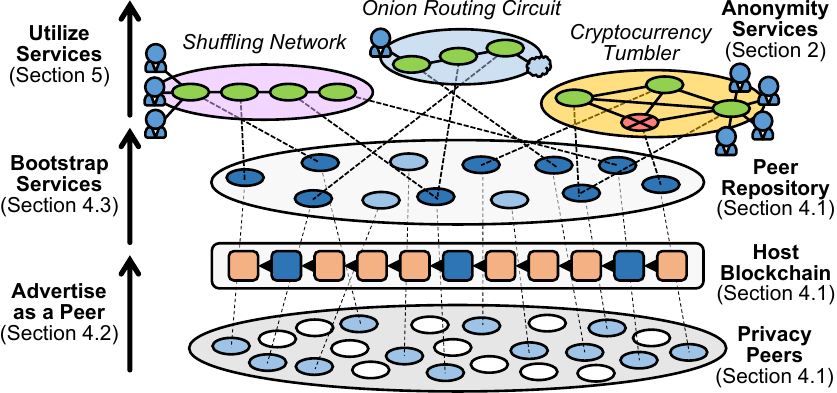}
    \Description[\name securely bootstraps distributed anonymity services from a peer repository that is maintained by having peer operators advertising themselves periodically via a host blockchain.]{
    This figure shows on a high level how \name's securely bootstraps distributed anonymity services, such as shuffling networks, onion routing circuits, or cryptocurrency tumblers.
    Operators of privacy peers have to periodically advertise their willingness to contribute to an anonymity service that is to be bootstrapped on a host blockchain.
    Creating valid advertisements adds a privacy peer to a (Sybil-resistant) peer repository, which then serves as the basis for bootstrapping anonymity services.}
    \caption{
        High-level design overview of \name, our medium for securely bootstrapping anonymity services.
    }
    \label{fig:teaser}
\end{figure}

\begin{figure*}[t]
    \captionsetup[subfloat]{farskip=0pt,captionskip=3pt}
  \centering
  \subfloat[Anonymity Network]{
        \centering
        \vspace{-1em}
        \includegraphics{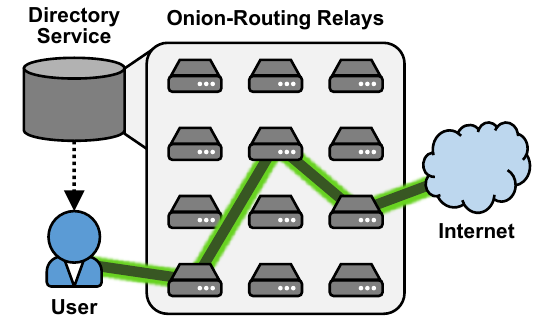}
        \label{fig:background:tor}
  }
  \rulesep
  \subfloat[Shuffling Network]{
        \centering
        \vspace{-1em}
        \includegraphics{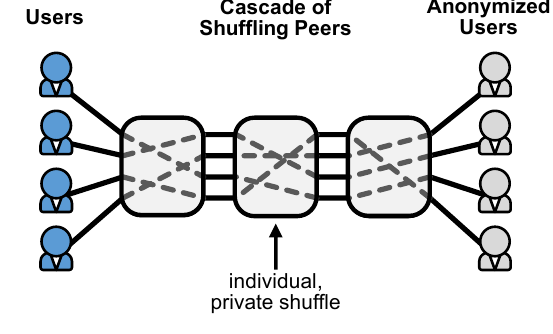}
        \label{fig:background:chaum}
  }
  \rulesep
  \subfloat[Cryptocurrency Tumbler]{
        \centering
        \includegraphics{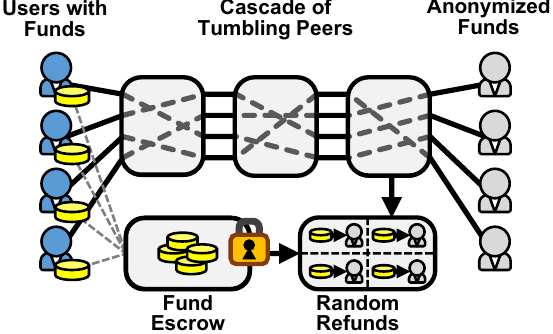}
        \label{fig:background:tumblers}
  }
  \Description[This figure illustrates the basic operation of currently deployed anonymity services.]{
      Figure~\ref{fig:background:tor} shows the basic operation of an anonymity network based on onion routing.
      The user obtains an index of available relays from a logically centralized directory service and creates a cascade of proxies from randomly selected relays, \ie the circuit.
      All user traffic subsequently traverses the circuit using layered encryption to prevent any party from linking the user to her traffic.

      Figure~\ref{fig:background:chaum} shows the operation of a shuffling network.
      Multiple users fix a cascade of shuffling peers and submit messages to the first node of the cascade, again using layered encryption.
      Each shuffling peer randomly permutes its obtained list of encrypted messages and forwards this list to the next shuffling peer, who does the same.
      Finally, the last shuffling peer lifts the last encryption layer and can release all messages, which are now unlinked from their originators.

      Figure~\ref{fig:background:tumblers} show how cryptocurrency tumblers extend the notion of a shuffling network to unlink users' digital coins from their identity.
      Before shuffling the associations between users and their funds, each user transfers her coins to an escrow address, which is jointly controlled by the cascade of tumbling peers.
      After the shuffling took place, the tumbling peers decrypted a set of fresh destinations to send the escrowed funds to.
      This process relies on threshold cryptography to prevent single tumbling peers from, \eg stealing escrowed funds.
  }
\caption{
	Well-known distributed anonymity services encompass (a) anonymity networks, such as Tor, for anonymous Internet communication, (b) message-shuffling networks, and (c) cryptocurrency tumblers to increase users' financial privacy.
    \vspace{-0.7em}
}
\end{figure*}

In this paper, we propose to outsource privacy-enhancing tasks to small networks of peers selected randomly in a secure, unbiased, and transparent fashion from a Sybil-resistant peer repository.
We introduce \emph{\name} as a medium for indexing and bootstrapping these anonymity services on top of a public host blockchain, which provides accepted means to maintain an immutable and transparent event log.
As we illustrate in Figure~\ref{fig:teaser}, peers join by periodically publishing advertisements containing a small proof of work (PoW) to the host blockchain.
Peer operators thus need to periodically invest hardware resources into refreshing their membership within a limited time frame, and all participants can locally derive \name's state by monitoring the host blockchain.
Hence, \name creates a Sybil-resistant index of privacy peers from which users can then request bootstrapping new anonymity services.
Users can choose privacy peers or established anonymity services from this index to cater to their individual privacy requirements.
We exemplarily build \name on top of Bitcoin to showcase its low requirements as well as the small storage footprint it has on its host blockchain, and to show that our system does not require sophisticated blockchain features, such as smart contracts, to operate.

\pagebreak
\subsubsection*{\bfseries Contributions}
\begin{itemize}
    \item{By analyzing existing anonymity services (Section~\ref{sec:background}), we identify a lack of secure bootstrapping for such services (Section~\ref{sec:scenario}).}
    \item{Through \name\footnote{Python-based implementation available at: \url{https://github.com/COMSYS/anonboot}}, we show that public blockchains are a suitable basis to create such a secure bootstrapping process (Section~\ref{sec:design}) for heterogeneous established use cases (Section~\ref{sec:use-cases}).}
    \item{We show that PoW and peer election can prevent adversaries from gaining advantages over honest peer operators (Section~\ref{sec:security}).}
    \item{\name scales to repositories of, \eg \num{1000} privacy peers and large user bases with only low storage impact on its host blockchain and low, tunable costs for its participants (Section~\ref{sec:eval}).}
\end{itemize}

\section{Available Anonymity Services}
\label{sec:background}

We identify three categories of distributed anonymity services for outsourcing privacy management: Internet anonymity networks, message shuffling networks, and cryptocurrency tumblers.

\subsection{Anonymity Networks}
\label{sec:background:tor}

\emph{Anonymity networks}, such as Tor~\cite{2004_dingledine_tor}, enable low-latency and anonymous Internet communication through onion routing, \ie tunneling users' traffic through a user-selected circuit under a layered encryption, as we exemplify in Figure~\ref{fig:background:tor}.
The user creates her circuits locally at random, but she also considers performance metrics, such as available bandwidth at individual nodes~\cite{2017_tor_path_spec}, as well as node-specific policies, \eg exit nodes only performing requests to certain ports on the user's behalf~\cite{2004_dingledine_tor}.
Tor provides the information required to build circuits through a \emph{directory} that is maintained by exceptionally trusted \emph{directory servers}~\cite{2004_dingledine_tor}.
These, currently ten~\cite{2009_tor_metrics_authorities}, directory servers are vetted by the Tor project maintainers, and users must trust that those directory servers do not collude~\cite{2009_panchenko_nisan}.
To further increase the reliability of this directory, relays are actively being probed~\cite{2014_winter_spoiled_tor,2011_chakravarty_tor_snooping}, and users can report misbehavior to the Tor project~\cite{2014_tor_report}.
Thus, misbehaving nodes are flagged in the directory to enable users to avoid such relays~\cite{2014_tor_report}.

\vspace{-0.7em}
\subsubsection*{\bfseries Takeaway}
Tor relies on an index of available nodes and their properties but requires trusted authorities to maintain this index.

\subsection{Message Shuffling Networks}
\label{sec:background:chaum}

Long before the recent proliferation of anonymity networks, David Chaum introduced networks for oblivious message shuffling~\cite{1981_chaum_mixnet}, to which we refer to as \emph{shuffling networks}, as a means to realize anonymous mail systems that provide sender anonymity, \eg to protect whistleblowers from retribution.
Figure~\ref{fig:background:chaum} showcases the basic user interaction with such a shuffling network.
Similarly to anonymity networks, users relay their messages through a cascade of known \emph{shufflers}, again after encrypting them in layers.
However, \emph{multiple} users shuffle their messages through the \emph{same} cascade of nodes to achieve a vastly reduced overhead.
These shufflers hence, one after another, receive the batch of encrypted messages of which they can lift only the outermost encryption layer.
After decrypting the message batch, each shuffler obliviously shuffles the batch's messages and forwards the result to the subsequent shuffler.
Therefore, shufflers are unable to correlate other shufflers' input and output batches.
As long as one shuffler remains honest, no passive adversary can deanonymize the users from now on.

However, shuffling networks are often prone to active attacks, such as denial of service (DoS) or replacing encrypted messages~\cite{2003_danezis_mixminion}.
Furthermore, adversaries can easily operate full shuffling networks at low costs since those networks are fixed and small in size.

\vspace{-0.7em}
\subsubsection*{\bfseries Takeaway}
Users need to trust that non-colluding operators run the shuffling network faithfully, which is especially challenging due to the current lack of a widely accepted index of shuffling networks.

\subsection{Cryptocurrency Tumblers}%
\label{sec:background:tumblers}

Multiple analyses of public blockchains, especially Bitcoin~\cite{2013_reid_anonymity_analysis,2013_katzenbeisser_anonymity_analysis,2013_meiklejohn_anonymity_analysis}, debunked the initial hope that cryptocurrencies provide sufficient user privacy by not building up long-lived identities~\cite{2008_nakamoto_bitcoin}.
To counteract curious blockchain observers, \emph{cryptocurrency tumblers}, or \emph{cryptotumblers}, break the linkability of privacy-aware users and their funds.
Cryptotumblers pool the funds of multiple users and then pay out random coins of the same value to each user such that the new coin owners are unknown to blockchain observers.

Cryptotumblers evolved over time, yielding different generations and flavors to appropriately address users' security and privacy concerns.
First, users of centralized cryptotumblers require strong trust in the service operator to not steal their funds or disclose the shuffling history at a later point to deanonymize users.
Series of alleged scams~\cite{2018_ziegeldorf_coinpartyv2,2014_badbitcoin_scams,2015_vasek_scams}, however, underpin the need for further \emph{technical protection}, \eg holding the cryptotumbler accountable~\cite{2014_bonneau_mixcoin}.

The first generation of distributed cryptotumblers let privacy-aware users jointly simulate a centralized tumbler by creating one large transaction with unlinkable inputs and outputs~\cite{2013_maxwell_coinjoin,2014_ruffing_coinshuffle}.
As the mixing is only performed if all users agree on the transaction's correctness, this approach is much more secure than involving a trusted third party.
However, single users can stall the mixing, which the other users must be able to detect to re-run the mixing without the misbehaving user~\cite{2014_ruffing_coinshuffle}.
Another branch of cryptotumblers aims for providing a distributed mixing service~\cite{2015_ziegeldorf_coinparty,2018_ziegeldorf_coinpartyv2,2018_meiklejohn_moebius}, \ie mix users' funds on their behalves without the risks involved with centralization.
While Möbius~\cite{2018_meiklejohn_moebius} achieves this via an Ethereum smart contract, CoinParty~\cite{2015_ziegeldorf_coinparty,2018_ziegeldorf_coinpartyv2} implements a blockchain-external service via a shuffling network and secure multiparty computation (SMC), and thus can also be used for mixing cryptocurrencies without support for smart contracts, \eg Bitcoin.
In Figure~\ref{fig:background:tumblers}, we illustrate the operation of such a CoinParty-like distributed cryptotumbler.
Using threshold signatures among the mixing peers prevents single adversaries from stealing funds, and secret-shared checksums are used to hold misbehaving mixing peers accountable during CoinParty's shuffling phase~\cite{2018_ziegeldorf_coinpartyv2}, \eg if attempting to perform attacks known from shuffling networks (\cf Section~\ref{sec:background:chaum}). 
However, this additional protection can only tolerate adversaries controlling a share $f_S\!<\!1\slash3$ of the service's privacy peers due to the application of SMC~\cite{1988_benor_smc_completeness}.

\vspace{-0.7em}
\subsubsection*{\bfseries Takeaway}
Although distributed cryptotumblers can increase the user's privacy, they either rely on smart contracts or are prone to Sybil attacks, \ie single adversaries mimicking a distributed service.
To the best of our knowledge, providing a technical medium to securely bootstrap cryptotumblers is still an open problem~\cite{2014_ruffing_coinshuffle}.

\begin{figure*}[t]
    \centering
    \includegraphics{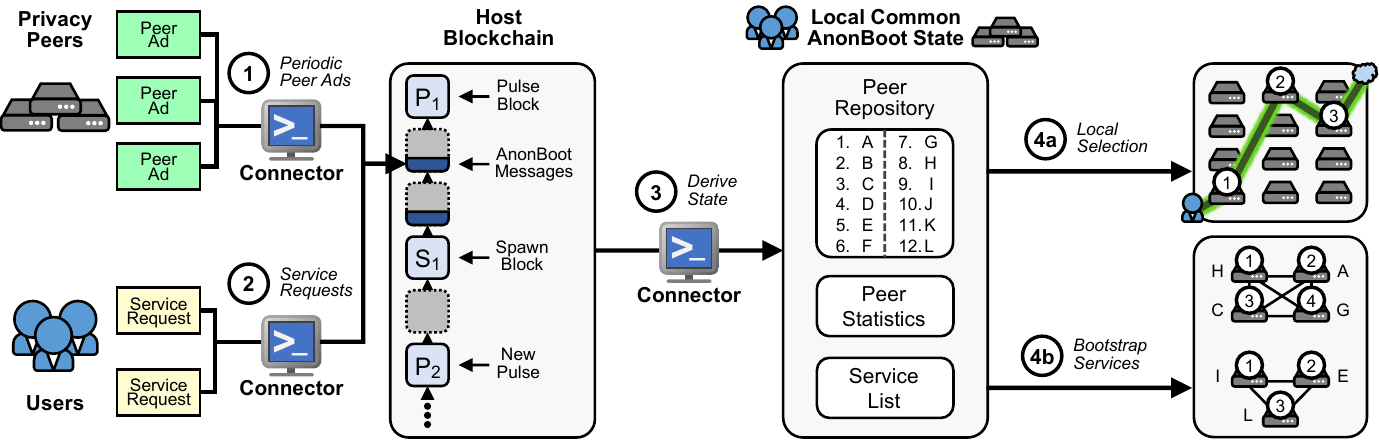}
    \Description[\name works in four phases that are orchestrated by each user's connector, which interacts with the host blockchain.
    Privacy peers first issue their peer advertisements periodically before users may request anonymity services to be bootstrapped.
    Once the participants locally derived the updated state, either users may locally select privacy peers directly, or privacy peers can continue to bootstrap requested services based on this state.]{
    This figure shows the detailed design overview of \name, which operates in four recurring steps.
    All interaction with the host blockchain is orchestrated by a connector, \ie \name's client software.
    In Step~1, privacy peers willing to join the peer repository have to create peer advertisements and store those on the host blockchain.
    In Step~2, users may issue service requests on the host blockchain, which will subsequently trigger bootstrapping a new anonymity service.
    All activity during Steps 1 and 2 is initiated by the next pulse block being mined on the host blockchain, \eg every $n$-th block may be interpreted as a pulse block by \name.
    Similarly, after the pulse block a spawn block will be mined, which closes the current pulse for writing new peer advertisements or service requests.
    After the spawn block has been mined, in Step~3, all participants locally derive the same \name state from the last pulse's on-chain messages.
    The \name state engulfs the peer repository of successfully advertised privacy peers, statistics about privacy peers, and the list of currently active anonymity services.
    In Step~4, user's can directly select privacy peers from the peer repository, \eg to establish a circuit for an anonymity network.
    Furthermore, privacy peers process the pulse's service request locally and find out whether they have been elected to bootstrap a new anonymity service.
    }
    \caption{
        In \name, peers periodically advertise themselves on the host blockchain while solving small PoW puzzles to prevent Sybil attacks.
        Users can bootstrap different anonymity services, which are then utilized independently of \name.
    }
    \label{fig:design:overview}
    \vspace{-0.5em}
\end{figure*}

\section{Scenario and Design Goals}%
\label{sec:scenario}

Based on existing anonymity services and the lack of proper bootstrapping processes, we now specify our scenario and design goals.

\subsection{A Generalization of Anonymity Services}
\label{sec:scenario:model}

As we discussed in Section~\ref{sec:background}, technical means for securely bootstrapping distributed anonymity services are currently lacking.
For a holistic solution resolving this lack of means to establish trust, we derive our scenario from the diverse landscape of existing services.

We assume a group of \emph{privacy-aware users} who seek to utilize an anonymity service that increases their privacy on their behalf.
To provide sufficient security and privacy guarantees, the users require that multiple independent operators of \emph{privacy peers} jointly offer distributed anonymity services.
Due to only limited scalability of network sizes of existing anonymity services, we assume that only a few privacy peers (\eg $<\!100$) provide services to much larger user groups.
Service provision is thus prone to Sybil attacks.

To account for local user decisions, such as creating Tor circuits (\cf Section~\ref{sec:background:tor}) or a minimum number of independent peers jointly providing a service, the user needs means to \emph{securely discover} available peers and already established anonymity services.
Furthermore, she has to \emph{establish trust} in the faithful setup of those services even if she does now know the peer operators.
Finally, the service discovery must allow for pooling users' anonymization efforts, as is required for shuffling networks or cryptotumblers.
Additionally, we need to \emph{incentivize} maintaining an honest majority of privacy peers.
However, we assume that a share of privacy peers will still act maliciously and aim to, \eg deanonymize users, stall the service, or inflict other damages such as theft through cryptotumblers.

In conclusion, users need to be ensured that they only utilize distributed anonymity services that act faithfully, \ie the majority of the respective peers are honest.
However, especially the setup and discovery of such services currently constitute weak points that adversaries could exploit to infiltrate anonymity services.

\subsection{Design Goals for Secure Bootstrapping}
\label{sec:scenario:goals}

The goal of our work is to create a decentralized medium for bootstrapping distributed anonymity services in a trustworthy manner and allowing privacy-aware users to discover both available peers and anonymity services.
To achieve this goal, we identify the following main requirements and features.

\vspace{-.7em}
\refstepcounter{goal}
\subsubsection*{\bfseries \gref{goal:bootstrapping}~Trustworthy Bootstrapping}\label{goal:bootstrapping}
Our medium must provide technical means to establish trust in available anonymity services and hence must be trustworthy itself.
To this end, a decentralized design is reasonable to eliminate the need for users' trust in any dedicated medium operator.
Furthermore, the medium must mitigate Sybil attacks to prevent its infiltration through adversaries.
Finally, the medium must still remain in control over the setup of offered anonymity services through a secure bootstrapping procedure.

\vspace{-.7em}
\refstepcounter{goal}
\subsubsection*{\bfseries \gref{goal:discovery}~Secure and Lightweight Service Discovery}\label{goal:discovery}
Our medium must only relay users to privacy peers and services that have been bootstrapped in a trustworthy manner.
Previous approaches have proposed piggybacking node discovery for peer-to-peer systems onto a well-established decentralized medium such as IRC~\cite{2007_knoll_bootstrapping}.
For such approaches, service discovery must limit its impact on the host system to facilitate the adoption of the bootstrapping process.

\vspace{-.7em}
\refstepcounter{goal}
\subsubsection*{\bfseries \gref{goal:applicability}~Broad Applicability}\label{goal:applicability}
In Section~\ref{sec:background}, we discussed the variety of existing anonymity services.
Consequently, we must account for this variety and allow users to discover and utilize different services for diverse applications.
Finally, users should be able to use anonymity services corresponding to their individual preferences.

\vspace{-.7em}
\refstepcounter{goal}
\subsubsection*{\bfseries \gref{goal:scalability}~Scalability.}\label{goal:scalability}
Sufficiently large user bases are crucial to achieving high privacy levels via anonymity services.
Our medium must thus effortlessly scale to large numbers of users and privacy peers.

\vspace{-.7em}
\refstepcounter{goal}
\subsubsection*{\bfseries \gref{goal:incentives}~Operator Incentives.}\label{goal:incentives}
Current honest anonymity services are typically offered on a voluntary basis~\cite{2004_dingledine_tor}.
However, if the effort of signaling honesty through our medium to publicly offer anonymity services becomes burdensome for operators, the number of volunteers might decrease.
Hence, our medium must also consider the option to compensate for operators' efforts in its design.

\section{\name: A Medium for Securely Bootstrapping Anonymity Services}
\label{sec:design}

In this section, we first provide an overview of \emph{\name} and then describe in detail how \name maintains a Sybil-resistant peer repository on top of a public host blockchain through standard transactions.
Finally, we elaborate on how \name bootstraps anonymity services from this repository, \ie how we elect privacy peers and then hand over control to the elected peers.

\subsection{Design Overview}
\label{sec:design:overview}

The main goal of \name is to provide a medium for \emph{securely} bootstrapping distributed anonymity services that typically consist of only a few \emph{privacy peers}.
\name maintains a robust distributed \emph{state} of available privacy peers and bootstrapping requests without storing privacy-compromising information.
To this end, \name relies on the immutable ledger of a public \emph{host blockchain} as the current state-of-the-art medium for communication and consensus without the need to rely on special trust in particular peers.
By having privacy peers periodically \emph{advertise} themselves on-chain through proof of work (PoW), \name maintains a \emph{Sybil-resistant peer repository}.
This way, the peer repository only contains recent privacy peers and adversaries need to invest resources in \emph{maintaining} their influence rather than increasing it over time.
\name realizes trustworthy bootstrapping \gref{goal:bootstrapping} by dynamically \emph{electing} privacy peers based on these advertisements and further on-chain entropy.
Thereby, \name prevents adversaries from manipulating peer election to gain an advantage over honest operators.

All participants locally operate a \emph{connector} for all interactions with the host blockchain.
The connector publishes new messages to the host blockchain and monitors it for new events.
Based on these events, the connector updates \name's state.
In this work, we detail how Bitcoin can be used as \name's host blockchain despite its very restricted intended ways to insert application-level data to show that \name can operate on top of most blockchains.
Furthermore, the Bitcoin network is well-established with around \num{10000} reachable nodes~\cite{2013_yeow_bitnodes} vetting its blockchain and thus providing a strong trust anchor regarding \name's privacy peers.

In Figure~\ref{fig:design:overview}, we present an overview of \name's bootstrapping process in four steps:
First, in \step{1}, privacy peers \emph{advertise} themselves on the host blockchain.
Subsequently, in \step{2}, users \emph{request} bootstrapping new anonymity services from a random set of advertised privacy peers.
Next, in \step{3}, all participants locally derive a common \name state.
Finally, based on their state, users can either (\step[a]{4}) \emph{locally select} privacy peers for personal anonymity services without the need for full synchronization, or (\step[b]{4}) privacy peers \emph{bootstrap} a new shared anonymity service.
We now provide a more detailed overview of these individual steps.

\vspace{-.7em}
\subsubsection*{\bfseries Periodic Peer Ads}
In \step{1}, \name creates a Sybil-resistant \emph{peer repository} by requiring privacy peers interested in providing anonymity services to periodically issue \emph{advertisements} on the host blockchain.
Peer operators need to periodically refresh their advertisements at the start of each refreshment period, or \emph{pulse}, while solving a small \emph{PoW puzzle}.
This core element of \name establishes a Sybil-resistant \emph{peer repository} as peer operators need to invest their hardware resources at the start of each pulse to remain in the peer repository.
To mitigate the advantage adversaries may gain through dedicated mining hardware, the exact design of the PoW puzzles is a crucial parameter of \name (\cf Section~\ref{sec:security:pow}).

\vspace{-.7em}
\subsubsection*{\bfseries Service Requests}
In \step{2}, privacy-aware users may issue aggregatable on-chain \emph{service requests} to request bootstrapping a shared anonymity service, \eg a shuffling network or a cryptotumbler, after a fixed-length negotiation phase.
Service requests specify the type of the anonymity service as well as service-specific parameters such as minimum required sizes of anonymity sets.

\vspace{-.7em}
\subsubsection*{\bfseries Derive State}
In \step{3}, all participants locally process advertisements and service requests from the host blockchain to derive and verify \name's current state.
By locally processing all on-chain service requests, all participants maintain a common \emph{service list}.
For processing requests, \name defines a randomized \emph{peer election} that is inspired by blockchain sharding~\cite{2018_kokoris_omniledger,2016_luu_elastico,2018_zamani_rapidchain} to select random subsets of compatible privacy peers, which then jointly provide the requested service.
Peer election is based on a pseudo-random number generator that is seeded with tamper-resistant entropy drawn from the host blockchain to enable all participants to locally derive the same service list.
After discarding invalid or delayed messages, all participants obtain the same state, \ie the current peer repository and statistics about previously discovered peers.

\vspace{-.7em}
\subsubsection*{\bfseries Local Selection \& Service Bootstrapping}
\step{4} finalizes the bootstrapping process provided by \name with two possible actions for users.
Either, users directly perform an instant \emph{local peer selection} only based on the peer repository (\step[a]{4}), \eg to establish a Tor circuit.
Alternatively, users browse the service list (\step[b]{4}) for securely bootstrapped anonymity services~\gref{goal:discovery}.
Since all privacy peers derive the same state as users, they can check whether they were elected to provide a shared anonymity service and subsequently bootstrap these services by contacting other elected privacy peers.
In both cases, communication is initiated through the \name connector, which then hands over the control entirely to the underlying anonymity protocol.

Our design ensures that \name only indexes anonymity services that are created in a Sybil-resistant manner as long as the peer repository itself consists of an honest majority.
The maintenance of an honest peer repository is therefore essential to \name's security.
To increase the willingness of honest peer providers to participate, \name can further decrease peer operators' costs by increasing the pulse duration, or existing anonymity services can be augmented with means for financial compensation, \eg via anonymous micropayments~\cite{2017_green_bolt}.
We argue that the increased robustness against adversaries offered by \name is worthwhile for privacy-aware users even if they are required to compensate privacy peer operator's costs.
However, \name is explicitly also operable by volunteers as long as its periodicity is tuned to prevent their recurring costs from becoming prohibitively high.
In the following, we present \name's protocol design in more detail.

\begin{figure}[t]
	\centering
	\input{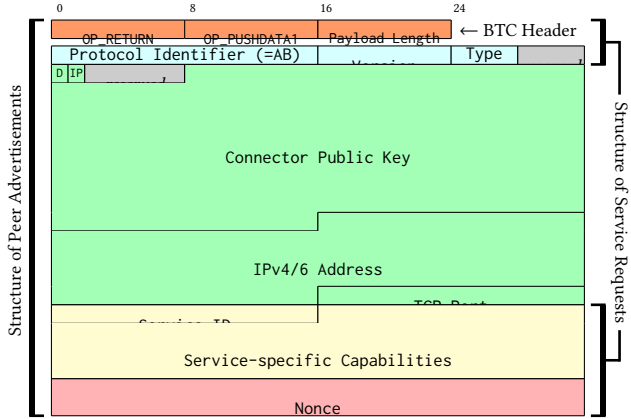}
    \Description[In \name, both peer advertisements and service requests share a similar structure utilizing Bitcoin's OP\_RETURN transactions.]{
    This figure shows the generalized message layout used in \name when operating on top of Bitcoin.
    Both peer advertisements and service requests share the same message structure based on Bitcoin's OP\_RETURN transactions.
    Both message types contain a protocol identifier for \name, the used protocol version, and message type.
    Peer advertisements also communicate its connector's public key as well as contact information in the form of IP address and TCP port for subsequent interaction.
    Further, both message types communicate service-specific capabilities, albeit with different semantics.
    In peer advertisements, the capabilities denote capabilities a privacy peer offers, whereas users specify required capabilities in their service requests.
    Finally, all messages contain a nonce.
    This nonce serves as the solution to the proof of work required for valid peer advertisements and as further entropy for peer election in service requests.
    }
	\caption{
        \name can run on top of Bitcoin using \code{OP\_RETURN} transactions.
        Peer advertisements convey peers' contact information, capabilities, and the required PoW.
        Service requests bootstrap a service based on the capabilities and using the nonce as further entropy for peer election.
	}
	\label{fig:design:advertisements}
    \vspace{-1em}
\end{figure}

\begin{figure*}[t]
    \centering
    \includegraphics{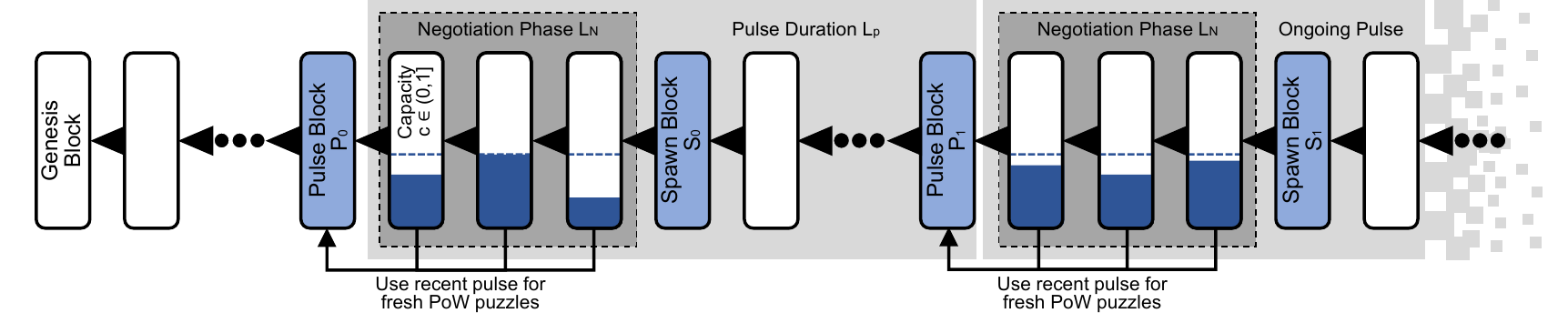}
    \Description[\name operates in pulses, which begin with a negotiation phase, during which participants may issue peer advertisements and service requests.
    This negotiation phase is concluded once the spawn block has been mined.]{
    This figure shows how \name operates in pulses of duration $L_P$, \ie one pulse covers $L_P$-many blocks on the host blockchain.
    During the negotiation phase of length $L_N$ at the start of each pulse, participants may issue peer advertisements and service requests up to a fixed capacity $c$ of the blocks' maximum size.
    Peer advertisements incorporate the pulse block into solving their PoW puzzle to ensure freshness of the advertisement.
    The first block being mined after the negotiation phase is further used to seed subsequent peer election and thus spawn new services, hence we refer to it as a spawn block.
    }
    \caption{
        Peer advertisements are written to the host blockchain and must be renewed by the \name peers after each pulse, incorporating PoW over the most recent pulse block to ensure freshness.
        Only peer advertisements published during the negotiation phase are considered valid, where miners are advised to optionally not exceed the desired capacity of \name messages per block.
        New anonymity services are bootstrapped after the negotiation phase based on the next spawn block.
        \vspace{-0.8em}
    }
    \label{fig:design:pulses}
\end{figure*}

\subsection{Sybil-Resistant Index of Peers and Services}
\label{sec:design:advertisements}

\name relies on its host blockchain to maintain its Sybil-resistant peer repository and to instantiate new anonymity services based on users' requests.
We now detail how \name can use Bitcoin as its host blockchain, only relying on standard transactions.
All concepts carry over to other blockchains, especially to systems that can process arbitrary messages on-chain through smart contracts, \eg Ethereum.
After presenting the basic structure of \name's Bitcoin-compatible messages, we thoroughly describe the message layout of peer advertisements and service requests.
Finally, we further elaborate on how \name enforces periodically refreshing messages to remain Sybil-resistant as well as fair toward honest privacy peers, and how it reduces its impact on the host blockchain.

\vspace{-0.2em}
\subsubsection*{\bfseries Basic Message Layout}

In Figure~\ref{fig:design:advertisements}, we show the generalized structure of a Bitcoin-compatible \name message.
Those messages are either \emph{peer advertisements} or \emph{service requests}.
All messages consist of \code{OP\_RETURN} Bitcoin transactions, which are allowed to carry up to \SI{80}{\byte} of payload data~\cite{2018_matzutt_contents}.
This structure results in an unavoidable \SI{3}{\byte}-long Bitcoin header consisting of the \code{OP\_RETURN} operation and the payload's length~\cite{2010_bitcoinwiki_script}.
The following \name header contains a protocol identifier (\code{AB}), as is common for \code{OP\_RETURN}-based protocols~\cite{2017_bartoletti_opreturn}, as well as the protocol version and message type.
For extensibility reasons, we reserve four bits for future use.

\vspace{-0.2em}
\subsubsection*{\bfseries Peer Advertisements}

Privacy peers join \name's peer repository by periodically refreshing and publishing \emph{peer advertisements} to the host blockchain.
As we detailed in Figure~\ref{fig:design:advertisements}, peer advertisements convey three main pieces of information for users and other privacy peers: (a)~the peer's \emph{contact information}, (b)~its \emph{capabilities}, and (c)~a \emph{solution} of its PoW puzzle.
While sharing their capabilities and contact information is required for coordinating the peer election (Section~\ref{sec:design:spawning}), ensuring Sybil resistance via peer advertisements is crucial for \name's promised security properties.

First, each privacy peer announces the \emph{contact information} of its connector so that users and other privacy peers can contact it securely in the following.
The privacy peer announces its connector's public key as well as a pair of IP address and port for incoming connections.
This indirection through a connector enables a unified connection interface for all anonymity services supported by \name.
However, if the advertised service's required contact information fits into the peer advertisement, the privacy peer may set the \code{D}-flag to indicate the direct reachability of the service, \ie the connector can be bypassed.
By setting the \code{IP}-flag, the privacy peer toggles whether it is reachable via IPv4 or IPv6, respectively.
Similarly, we reserved six additional bits for future use to remain flexible regarding other formats of contact information.

Second, each peer advertises its \emph{capabilities}.
These capabilities consist of a service identifier denoting which anonymity service the privacy peer supports as well as service-specific capabilities.
This design supports the integration of a diverse landscape of anonymity services (\cf Section~\ref{sec:background}) as well as future services into \name and thus respects our requirement for broad applicability~\gref{goal:applicability}.
These service-specific capabilities help users request services or locally select privacy peers that suit their individual needs.
While smart contract-based host blockchains can process arbitrary messages and thereby enable the fine-grained expression of privacy peers' capabilities, the space limitations of Bitcoin's \code{OP\_RETURN} payloads restrict this expressiveness.
For instance, creating Tor circuits relies on potentially complex relay descriptors~\cite{2007_tor_dir_spec} that easily exceed available space and otherwise would impose a large overhead on the host blockchain.
We make \name operable even in such restricted environments by allowing privacy peers to advertise coarse-grained capabilities as a browsing aid that is subsequently verified and refined via the participants' connectors.

Finally, privacy peers need to include a small PoW in their peer advertisements to \emph{thwart Sybil attacks}.
To be effective, the PoW must be cryptographically tied to the peer's identity as well as a recent point in the host blockchain to prevent an adversary from pre-computing or reusing peer advertisements.
Only then, the PoW puzzle ensures that no peer can create disproportional numbers of peer advertisements compared to its hardware resources.

\vspace{-0.5em}
\subsubsection*{\bfseries Service Requests}

Users issue \emph{service requests} to express that they want \name to bootstrap a new anonymity service corresponding to their requirements.
Service requests closely resemble peer advertisements in their structure (\cf Figure~\ref{fig:design:advertisements}), but they do not contain contact information.
Further, the remaining fields are interpreted slightly differently.
Through the capabilities, users express what service they intend to use as well as minimum requirements for the service to be bootstrapped.
\name only allows users to request distinct classes of services through the capabilities to prevent a highly fragmented service list.
In contrast to privacy peers, users do not solve a PoW puzzle in their service requests.
Instead, users choose a random nonce, which \name will incorporate into its peer election to subsequently bootstrap the requested services.
This way, users can further thwart attempts by adversaries to interfere with the peer election.
A single service request will cause \name to instantiate the requested service to be used by an arbitrary number of users.
Hence, \name easily scales to large user bases~\gref{goal:scalability}.
However, users questioning the existing requests' randomness can issue redundant service requests and thus contribute to the entropy used for the peer election.
\name aggregates redundant requests or similar requests superseded by service requests with stronger requirements, \ie all requests' nonces influence the peer election, but only one service with the most restrictive capabilities of all aggregated service requests will be bootstrapped.
At this point, we leave defining strategies to simultaneously instantiate multiple similar services as future work.

\vspace{-0.5em}
\subsubsection*{\bfseries Pulse-based Message Release.}
\label{sec:design:advertisements:pulses}

In Figure~\ref{fig:design:pulses}, we illustrate \name's soft-state approach, which defines a \emph{pulse} of length $L_p$ in terms of block height on the host blockchain that triggers refreshing peer advertisements and accepts new service requests.
Every $p$ blocks, a new pulse starts and the most recent block on the blockchain serves as the new \emph{pulse block}.
Now all peers start to create their PoWs incorporating (a)~their connector's public key, (b)~a reference to the pulse block to ensure the freshness of advertisements, and (c)~a nonce solving the PoW.
To extract the maximum entropy from the pulse block, \name can apply extraction techniques, \eg as proposed by Bonneau \etal~\cite{2015_bonneau_beacon}.

For ideal fairness, all peers would have the same time window for providing a valid PoW.
However, \name must be able to cope with a potential backlog of valid peer advertisements since we have no means to reliably enforce prioritized consideration of \name messages after a pulse block was being mined.
Thus, we tolerate peer advertisements to be delayed throughout a \emph{negotiation phase} of length $L_N$ after each pulse.
This length should be chosen as short as possible to prevent a devaluation of PoWs provided by honest peers, but it simultaneously should allow for including all anticipated peer advertisements in time even if single miners deliberately ignore \name messages.
Furthermore, the negotiation phase provides some tolerance against accidental blockchain forks.
While peers must recompute their PoW if the host blockchain discards the pulse block, a fork does not require \name to skip an entire pulse.

The tunable duration of each pulse with its associated negotiation phase also allows to adjust the burden put on its participants as well as its host blockchain in a fine-grained manner and thus allows keeping the service discovery lightweight~\gref{goal:discovery}.
First, \name disincentivizes excessive creation of messages as honest peers will ignore all messages outside of a pulse's negotiation phase.
Second, increasing $L_p$ without changing $L_N$ reduces the number of messages required to maintain the peer repository, \ie costs for all peers are reduced, without weakening \name's Sybil resistance and only at the cost of the peer repository becoming less flexible.
However, \name still releases messages in bursts at the start of each pulse.
If these occasional message bursts prove to be burdening the host blockchain, \name-aware miners can follow an optional guideline to accept messages only up to a per-block \emph{capacity $c\in(0,1]$} without impacting \name negatively.
Furthermore, more awareness from miners on the host blockchain has the potential to further reduce costs of \name peers and thereby lower the bar for altruistic peer operators.
Either through updated consensus rules or novel, \name-tailored blockchain designs, miners can be incentivized to reserve up to $c\!\cdot\!100\%$ of their blocks during each negotiation phase for including \name messages at no costs.
For instance, full nodes may then reject blocks that ignore a current backlog of pending \name messages.
We further quantify how the host blockchain can steer the impact of \name in Section~\ref{sec:eval:footprint}.
While this approach requires that miners are not entirely oblivious of \name, it ensures that \name can operate at minimal costs without burdening the host blockchain.

\subsection{Bootstrapping Secure Anonymity Services}
\label{sec:design:spawning}

All privacy peers that regularly refresh their peer advertisements are eligible for providing anonymity services.
In this section, we describe how \name facilitates bootstrapping anonymity services based on the current pulse and its resulting peer repository.
After briefly describing how control is handed over from \name to its bootstrapped services, we consider users locally picking privacy peers directly from the peer repository and then provide details on how \name elects privacy peers to bootstrap publicly available, distributed anonymity services.

\vspace{-0.5em}
\subsubsection*{\bfseries Bootstrapping Users and Privacy Peers}
\name provides only a medium for establishing and finding trustworthy distributed anonymity services.
Its responsibility thus also involves enabling users to contact privacy peers that provide the requested anonymity service.
In most cases, peer advertisements will announce the contact information of the involved privacy peers' \name connector.
During the handover of control via her own connector, the user verifies the correctness of each peer's contact information, especially whether it possesses the private key corresponding to its advertisement.
If successful, the connectors perform a service-specific handover so that further interaction is now performed entirely according to the anonymity service protocol.
In cases where indirection through the connector is undesired, privacy peers may use the \code{D}-flag (\cf Section~\ref{sec:design:advertisements}) to signal that the contact information directly corresponds to the endpoint of its offered service.
However, a Bitcoin-backed \name only supports \code{OP\_RETURN}-based direct advertisements if they can hold all required contact information.

Depending on the particular anonymity service (\cf Section~\ref{sec:background}), users either contact (a)~only one privacy peer, (b)~all privacy peers of one anonymity service, or (c)~may only indirectly contact subsequent privacy peers for security reasons, \eg when establishing Tor circuits.
In cases where a direct connection to peers is prohibited, users can interleave bootstrapping with the anonymity service and incrementally contacting the new peers' connectors.
For instance, Tor builds circuits hop by hop~\cite{2019_tor_spec}, and thus users can contact the connectors of subsequent Tor nodes via partially established circuits, which aligns well with Tor's design~\cite{2019_tor_spec}.

\vspace{-0.5em}
\subsubsection*{\bfseries Local Selection of Peers}
The peer repository's Sybil resistance (\cf Section~\ref{sec:design:advertisements}) makes it a suitable replacement for centrally maintained directories.
Privacy-aware users individually monitor peer advertisements, which enable them to instantly select privacy peers based on their local view on the peer repository, \ie this peer selection is independent of \name's pulses.
Furthermore, users may base their decisions on individual security and privacy preferences, \eg they only select privacy peers who recently advertised themselves, or they may locally keep track of peer statistics, such as their first occurrence or how regularly they refresh advertisements.

When selecting privacy peers, the user verifies the correctness of those peers' advertisements and contacts their connectors.
To this end, users only have to passively monitor the host blockchain for valid peer advertisements from the current pulse.
Each peer that (a)~performed a valid and fresh PoW, (b)~is reachable via its connector's contact information, and (c)~advertised a valid corresponding public key is eligible to be selected by the user.
Ultimately, the user randomly selects a sample of peers she considers eligible replacing any inaccessible peers until the service can be provided correctly.

\begin{figure}[t]
    \centering
    \includegraphics{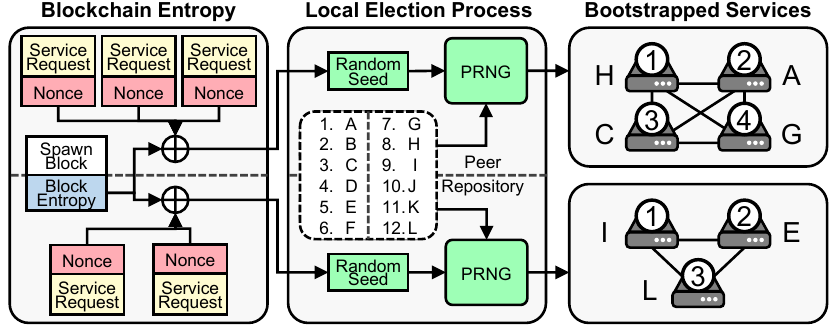}
    \Description[Peer election in \name is executed whenever a spawn block is being mined and the process is locally replayable as it involves seeding a pseudo-random number generator using only recent on-chain information.]{
        This figure shows how the \name connector locally executes the peer election process whenever a new spawn block has been mined.
        This process consists of pseudo-randomly draw a sample of privacy peers from the peer repository based on a random seed that is derived from recent on-chain information.
        This seed is derived by combining the nonces of all service requests that can be aggregated as well as entropy from the spawn block.
        Using this process, a list of privacy peers that shall jointly provide a requested anonymity service is drawn for each class of service requests in the current pulse.
    }
    \caption{
		A pseudo-random peer election based on blockchain entropy enables all participants to locally compute the same service lists from the peer repository.
    }
    \label{fig:design:bootstrap}
    \vspace{-0.7em}
\end{figure}

\vspace{-0.5em}
\subsubsection*{\bfseries Service Requests For Peer Election}
\name derives the demand for anonymity services from users' service requests during the negotiation phase (\cf Section~\ref{sec:design:advertisements}).
Based on these service requests, we must ensure that peers are chosen randomly in a transparent manner to provide a secure bootstrapping process.
We achieve this requirement through a locally replicable \emph{peer election} process that relies on a pseudo-random number generator (PRNG) and seeds derived from random values on the host blockchain.
This way, all participants obtain the same list of elected peers for each distinct service request for subsequent coordination.

In Figure~\ref{fig:design:bootstrap}, we present \name's peer election in more detail.
To derive the seed, we rely on two sources of entropy.
On the one hand, users submit \SI{8}{\byte}-long nonces with their service requests.
We aggregate the nonces of all matching service requests during one pulse, \ie requests for the same anonymity service with the compatible capabilities.
This approach allows a bootstrapping of anonymity services with a single service request for efficiency while it also offers privacy-aware users the chance to directly influence the peer election's randomness without spawning concurrent services that are potentially under-utilized.
On the other hand, we consider the \emph{spawn block} of each pulse, \ie the first block \emph{after} the pulse's negotiation phase has concluded.
Thus, an adversary cannot craft nonces to bias the peer election without mining the spawn block.
We incorporate entropy from this block into the seed for the PRNG to ensure its freshness.
All participants locally use the PRNG with this seed to elect peers for each service request and select a pseudo-random sample of privacy peers from the peer repository that is compatible with the service request.
A common ordering of the peer advertisements ensures that all participants select the same samples.
The peer election allows all participants to compute the same service list and thus synchronize in a decentralized manner.
Hence, \name helps users find the required entry points for using anonymity services, and the privacy peers learn whom to connect to when being elected to join a specific anonymity service.

\subsubsection*{\bfseries Conclusion of Design}
Our design of \name enables trustworthy bootstrapping~\gref{goal:bootstrapping} since (a)~it operates on top of a public host blockchain in a decentralized manner, (b)~it mitigates Sybil attacks through periodically refreshed and PoW-based peer advertisements, and (c)~it realizes a secure bootstrapping process using entropy from users as well as the host blockchain's mining process.
By exchanging messages through the host blockchain, our design facilitates secure service discovery with only a low impact on the host blockchain due to \name's parametrizable pulse length and per-block capacity~\gref{goal:discovery}.
Our protocol-agnostic message structure and handover of control moreover ensure a broad applicability of \name~\gref{goal:applicability}.
Finally, \name scales to large user bases as single service requests suffice to bootstrap anonymity services usable by arbitrarily many users~\gref{goal:scalability}.
In the following, we outline how to integrate different anonymity services into our medium and how \name can incentivize honest participation of privacy peers to satisfy the remaining design goal~\gref{goal:incentives}.

\section{Realizing Use Cases in \name}
\label{sec:use-cases}

After presenting the general medium provided by \name, we now discuss how the established anonymity services, which we presented in Section~\ref{sec:background}, can operate on top of this medium regarding the achievable benefits, the technical integration, and how to financially incentivize honest privacy peers' participation~\gref{goal:incentives}.
First, we discuss how \name's peer repository can realize a distributed directory service for anonymity networks.
Subsequently, we discuss the bootstrapping of shuffling networks and cryptotumblers, as both behave similarly in \name.

\subsection{Decentralized Onion Routing via \name}%
\label{sec:use-cases:tor}

\name's Sybil-resistant peer repository constitutes a cryptographically controlled replacement for otherwise logically centralized directory services.
Hence, our approach is beneficial if users expect operators to be corruptible or malicious.
Nevertheless, \name must allow users to still make informed choices about the establishment of circuits, and we must account for the infeasibility and insecurity of users directly contacting all peers of a circuit.

\subsubsection*{\bfseries Benefits}
In currently deployed anonymity networks, the directory service is essentially centralized.
For example, Tor is pre-shipped with a hard-coded list of currently ten \emph{directory authorities}~\cite{2009_tor_metrics_authorities}, which jointly maintain its directory~\cite{2007_tor_dir_spec}.
This approach leaves current anonymity networks vulnerable to viable attacks on the directory service~\cite{2004_dingledine_tor}.
Contrarily, \name allows creating a fully decentralized directory that is implicitly maintained through the host blockchain and locally verifiable by all \name participants.
Based on this directory, users can locally select privacy peers for their circuits as they currently do through Tor's directory service.

\subsubsection*{\bfseries Peer Advertisements}
Privacy peers can advertise themselves as onion routers.
However, Tor's directory service maintains extensive meta information about available peers~\cite{2007_tor_dir_spec}, which in most cases cannot be encoded in a single \code{OP\_RETURN}-based peer advertisement as required when \name shall operate on top of Bitcoin.
Among this meta information is the peer's contact information, cryptographic identity, available bandwidth, supported features, and exit policies, \ie access control list for connections to hosts on the public Internet~\cite{2007_tor_dir_spec}.
We thus make use of the peer advertisements' capabilities (\cf Section~\ref{sec:design:advertisements}) to encode an \emph{overview} of the peers' full meta information.
This overview is a coarse summary of a privacy peer's \emph{advertised} capabilities and should be indicative of its actual capabilities.
Users can then browse available privacy peers based on these advertised capabilities without additional delays.
When establishing a new circuit, the user should then request the chosen privacy peers' full server descriptors, verify that this descriptor matches the previously advertised capabilities, and check that the full descriptor is also compatible with the user's requirements.

\subsubsection*{\bfseries Bootstrapping Phase}
The circuits users establish within anonymity networks are intended to provide sender-receiver anonymity.
Hence, a critical constraint is that users only communicate directly with the first peer of a circuit.
\name naturally integrates with the resulting incremental circuit establishment of Tor~\cite{2019_tor_spec}:
The user incrementally establishes the next hop of her new circuit based on her selected peers' advertisements.
She contacts the new peer through her partially established circuit and attempts to hand over control to the Tor client through the connector.
If this handover of control fails, \eg due to an invalid advertisement, she terminates the connection to that peer and selects a replacement privacy peer.
Although an honest majority among privacy peers reduces the overhead of such security back-offs, enabling privacy peers to build up a positive reputation across consecutive peer advertisements promises to further reduce respective risks for users.

\subsubsection*{\bfseries Incentives}
If honest providers of onion routers must be compensated for investing their resources to periodically solve PoW puzzles and advertise themselves in \name, cryptocurrency-based service fees are a promising means for creating operator incentives.
However, on-chain payments bear high risks of implicitly recording information about users' circuits irrevocably.
We thus propose that users and privacy peers create anonymous unidirectional micropayment channels~\cite{2017_green_bolt}.
Although micropayment channels require an on-chain setup, users can protect their privacy due to the concurrent setup transactions of all users.
This way, users can pay peers who advertise themselves via \name for their service.

\subsection{Shuffling Networks and Cryptotumblers}%
\label{sec:use-cases:outsource}

\name's main advantage is to provide a medium for bootstrapping distributed anonymity services and to ensure their privacy peers' independence through its PoW puzzles and secure peer election.
Privacy-aware users thus gain the opportunity to rely on secure on-demand anonymization for, \eg message shuffling or increasing their financial privacy.

\vspace{-.2em}
\subsubsection*{\bfseries Benefits}
Distributed systems that outsource responsibility to a set of peers typically rely on secure multi-party computation (SMC)~\cite{2018_ziegeldorf_coinpartyv2,2017_ateniese_redactable}.
Unfortunately, scalability limitations of those SMC protocols hinder distributing responsibility among large sets of privacy peers.
Without carefully selecting the responsible privacy peers, insider adversaries thus can gain power and cause harm relatively easily.
However, our considered use cases of anonymous message disclosure and tumbling cryptocurrencies lack a trustworthy peer selection process, and adversaries are highly incentivized to attack such systems.
For example, an adversary could easily spawn numerous interconnected privacy peers, and thereby mimic a distributed cryptotumbler, tricking users into participation.
\name provides the ingredients to \emph{cryptographically ensure} through its Sybil-resistant peer repository and locally verifiable peer election that an adversary cannot bootstrap malicious services.
Hence, privacy-aware users reduce their individual risks when utilizing distributed anonymity services bootstrapped via \name.

\vspace{-.2em}
\subsubsection*{\bfseries Peer Advertisements}
The capabilities privacy peers need to advertise highly depend on the provided anonymity service.
Similarly to our previous use case, privacy peers should facilitate the users' browsability of anonymity services by advertising supported policies or security parameters.
However, \name does not consider the service-specific capabilities during its peer election but requires the service identifier used (\cf Section~\ref{sec:design:advertisements}) to ensure compatibility among privacy peers advertising the same service.

\vspace{-.2em}
\subsubsection*{\bfseries Bootstrapping Phase}
Privacy peers are partitioned by the identifier of the anonymity service they advertise to ensure compatibility during the bootstrapping process.
By locally replaying the peer election, each privacy peer gets to know (a)~whether it was elected to provide a service, (b)~which peers are elected to bootstrap the same service instance, and (c)~the peer's logical position within the new network.
Hence, privacy peers can independently configure and bootstrap the anonymity service.
Currently, we take a conservative approach and declare services stale after a couple of pulses to mitigate the impact of privacy peer churn and malicious services bootstrapped by chance.
However, conceptually, \name also supports bootstrapping long-lived anonymity services.

\vspace{-.2em}
\subsubsection*{\bfseries Incentives}
Since these use cases do not prohibit a direct connection between users and elected privacy peers, we can simplify our payment scheme proposed in Section~\ref{sec:use-cases:tor} and instead require users to pay an upfront fee (\eg as proposed by CoinParty~\cite{2018_ziegeldorf_coinpartyv2}).
We argue that the increased security provided by \name is worth compensating the privacy peer's efforts of solving PoW puzzles.

\vspace{-.2em}
\subsubsection*{\bfseries Takeaway}
In conclusion, \name provides a viable medium for bootstrapping anonymity services from a diverse set of available applications as it simultaneously mitigates malicious influences and compensates honest operators if privacy peers.

\section{Security Discussion}
\label{sec:security}

We assess \name's robustness against adversaries by discussing the implications of incorporating PoW into peer advertisements and arguing that active adversaries cannot bias the peer election.

\subsection{Proof of Work Against Sybil Attacks}
\label{sec:security:pow}

Requiring a PoW in each peer advertisement hampers an adversary's effort to control large portions of the peer repository and thus his overall influence.
However, the choice of the PoW scheme is paramount for \name's resilience against Sybil attacks.
We thus highlight the need for an appropriate PoW scheme but leave its final instantiation to be adapted to users' needs in future work.

Particularly, \name's PoW scheme must ensure that operators can only create peer advertisements at rates corresponding to their number of physical devices controlled while not excluding honest operators using commodity hardware.
While specialized hardware is known to provide huge advantages for CPU-bound PoW schemes such as Bitcoin's scheme, memory-bound PoW schemes such as Ethereum's Ethash~\cite{2014_wood_ethereum,2015_ethereum_ethash}, Cuckoo Cycle~\cite{2015_tromp_cuckoo}, Equihash~\cite{2017_biryukov_equihash}, or RandomX~\cite{2018_monero_randomx}, which was recently adopted by Monero~\cite{2019_shevchenko_randomx}, are promising candidates to be adapted for utilization with \name.
For instance, based on \code{openssl speed}, we observe that a server (two Intel Xeon Silver 4116, \SI{187.39}{\gibi\byte} RAM) outperforms a commodity desktop PC (Intel Core 2 Q9400 CPU, \SI{7.67}{\gibi\byte} RAM) by two orders of magnitude for Bitcoin's HASH256-based PoW scheme.
Further, Bitcoin mining hardware~\cite{2015_bitcoin_asics} reportedly outperforms our commodity PC by eight orders of magnitude, which clearly underlines the potential advantage of adversaries relying on specialized hardware to forge advertisements using CPU-bound PoW schemes.

Contrarily, initial measurements using Ethash (via \code{geth}'s CPU-based mining) and RandomX indicate that the same server only achieves a mere $7.5\times$ ($12.7\times$) speed-up over the desktop PC in terms of achievable hash rate using this PoW scheme.
Thus, relying on memory-hard PoW schemes is preferable to prevent adversaries with powerful devices or, \eg a botnet, from increasing their influence on the peer repository in an incommensurate manner~\cite{2017_biryukov_equihash}.

Finally, we address the challenge of steering the PoW puzzles' difficulty to account for improvements in hardware capabilities.
In contrast to cryptocurrency mining, \name's peer advertisements have no inherent concurrency, \ie the size of the peer repository does not influence the required difficulty for the PoW.
Assuming an honest majority, we can expect that privacy peers have an interest in keeping an appropriate PoW difficulty for security reasons.
Thus, we can dedicate unused bits in the peer advertisements (\cf Section~\ref{sec:design:advertisements}) to enable voting on increasing the difficulty.
Privacy peers would then update their local threshold for accepting the PoW in peer advertisements based on votes of the (honest) majority.

\subsubsection*{\bfseries Takeaway}
Utilizing a simple CPU-bound PoW scheme for our puzzles would significantly impact \name's security properties.
Contrarily, memory-bound PoW schemes constitute a secure building block to maintain a Sybil-resistant peer repository.
As for existing systems, such as Tor or Bitcoin, the reliability of \name's peer repository then depends on maintaining an honest majority, either on a voluntary basis or through operator incentives.
Finally, we can further leverage this honest majority to implement a self-regulated adaption of the puzzles' difficulty.

\begin{figure}[t]
    \centering
    \includegraphics{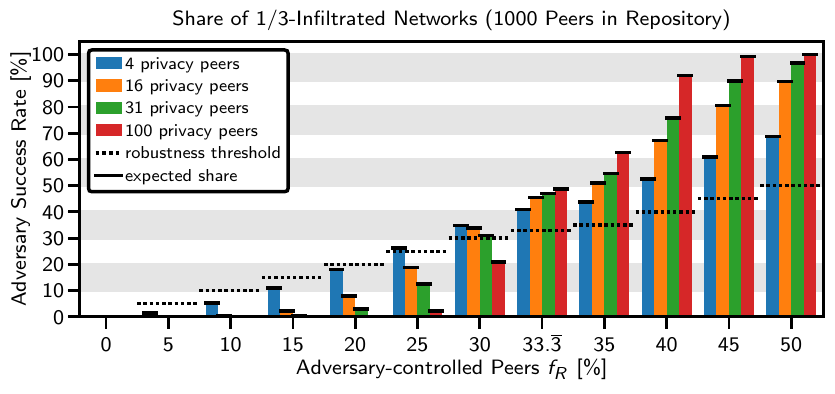}
    \Description[For varying network sizes of 4, 16, 31, and 100 out of 1000 peers, and increasing shares of overall privacy peers being controlled by an adversary, we repeatedly execute our peer election and analyze the adversary's success rate in controlling at least one third of the elected privacy peers.
    Our measurements show that \name's peer election remains robust against adversarial infiltration as long as the adversary does not control one third or more of the whole peer repository.]{
    This figure shows our results of empirically measuring an adversary's success rate of infiltrating anonymity services spawned through \name's peer election when controlling growing shares of a peer repository of size 1000.
    We increment the share of adversary-controlled privacy peers from 0\% to 50\% in steps of 5\% each and additionally consider the special case where the adversary controls one third of the peer repository.
    For each share, we spawn networks of sizes 4, 16, 31, and 100, respectively.
    Our experiment shows that peer election is in line with the expected success rate based on combinatoric considerations, and that our peer election remains robust as long as the adversary controls less than one third of the peer repository, with larger networks being more resilient against adversarial infiltration in this area.
    When controlling one third or more of the overall peer repository, the adversary has increasingly high success rates of infiltrating networks, rendering \name insecure to use assuming such high overall control by the adversary, as is the case for SMC-based anonymity services in general.
    }
    \caption{
        The success rate of an adversary to $\mathbf{1\!\slash\!3}$-infiltrate services by chance can be kept below the robustness threshold $\mathbf{T_{1\slash3}}$ for $\mathbf{f_R\!\le\!30\%}$, \ie peer election remains robust for relevant scenarios involving SMC-based anonymity services.
    }
    \label{fig:security:intruded-mixnets}
    \vspace{-0.7em}
\end{figure}

\vspace{-1.2em}
\subsection{Security of Bootstrapped Services}
\label{sec:security:selection}

The core design goal of \name is to securely bootstrap distributed anonymity services.
We have already shown that \name can maintain a Sybil-resistant peer repository, \ie adversaries cannot control a disproportional fraction of the peer repository.
However, adversaries can still enter the peer repository as long as they can create a valid PoW.
We now highlight that \name's peer election is robust against adversarial bias and that bootstrapped anonymity services can tolerate a share of adversarial privacy peers.

\vspace{-0.5em}
\subsubsection*{\bfseries Security of Local Peer Selection}

Anonymity services which rely on local peer selection only require the Sybil resistance provided by \name's peer repository (\cf Section~\ref{sec:security:pow}).
However, privacy peers are not always treated equally:
For example, in Tor, users change their first relays, \ie guard nodes, only infrequently~\cite{2017_tor_guard_spec} and they can only use exit nodes that support their requests~\cite{2004_dingledine_tor}.
By encoding the privacy peers' capabilities accordingly (\cf Section~\ref{sec:design:advertisements}), users can respect these properties when establishing circuits.
The peer repository hence constitutes a secure alternative to current directory services provided by trusted third parties.

\vspace{-0.5em}
\subsubsection*{\bfseries Robustness of Peer Election}

\name's peer election must properly protect its users, \ie bootstrap secure anonymity services.
To assess the influence of an adversary, we consider his chance of \emph{infiltrating} an anonymity service during peer election based on the share of privacy peers he controls.
An adversary successfully infiltrates an anonymity service if he controls a share $f_S\!\ge\!t_I$ of that service's privacy peers, exceeding its \emph{infiltration threshold $t_I$}, \ie he can defy the service's underlying security guarantees.
\Eg a malicious adversary infiltrates any SMC-based anonymity service when controlling $f_S\!\ge\!1\slash3$ of the peers~\cite{1988_benor_smc_completeness}.
Under this notation, we consider peer election to be robust if adversaries cannot increase their chance of infiltrating services beyond their share $f_R$ of privacy peers in the peer repository.
More formally, assuming that no adversarial share of the peer repository exceeds a threshold $t_R$, we define a \emph{robustness measure $\mathcal{R}(t_I, t_R)=1\!-\!\Pr(f_S\!\ge\!t_I\ |\ f_R\!\le\!t_R)$}.
We further define that the peer election is \emph{robust} iff $\Pr(f_S\!\ge\!t_I\ |\ f_R\!\le\!t_R)\le t_R =: T_I$ holds, \ie $T_I\!=\!t_R$ can be interpreted as a \emph{robustness threshold} against $t_I$-infiltration.
For instance, an adversary controlling up to $f_R\!\le\!10\%$ of the peer repository should only have a chance of $T_I\!=\!10\%$ to $t_I$-infiltrate an anonymity service.

Figure~\ref{fig:security:intruded-mixnets} highlights \name's robustness regarding SMC-based anonymity services, \ie $t_I\!=\!1\slash3$~\cite{1988_benor_smc_completeness}, which can tolerate up to $\left\lfloor n\slash 3\right\rfloor$ adversary-controlled privacy peers.
For desired networks consisting of \num{4}, \num{16}, \num{31}, and \num{100} peers (\ie $t_I\!=\!1, 5, 10, 33$) respectively, we measured the success of an adversary controlling a growing share $f_R$ of the peer repository to infiltrate anonymity services by chance due to our peer election.
To extract entropy from the pulse's spawn blocks, we rely on the Merkle tree root.
More secure entropy extraction can be achieved by applying more sophisticated randomness extractors~\cite{2015_bonneau_beacon}.
For our evaluation, we assume a peer repository consisting of \num{1000} peers, randomly elect peers for \num{100000} anonymity services for each scenario, and count the number of $1\slash3$-infiltrated services.
We also highlight the robustness threshold for comparison and provide the expected shares of $1\slash3$-infiltrated services based on combinatoric considerations.

Our evaluation reveals two major findings:
First, our peer election is \emph{fair} in that it almost perfectly yields the expected distribution of $1\slash3$-infiltrated services when electing honest and dishonest peers uniformly at random.
Second, the peer election remains robust as long as the adversary controls $f_R\!\le\!25\%$ of the peer repository.
For a growing power of the adversary, \name cannot guarantee robustness, although larger anonymity services yield better protection if the adversary controls a share of at most $f_R\!\le\!30\%$.
For all $f_R\!\ge\!1\slash3$, \name is not robust anymore as the adversary can infiltrate most SMC-based services.
However, in those cases, his control of the peer repository exceeds the infiltration threshold for SMC-based services; thus, we consider the peer repository insecure.

\name relies on entropy from the host blockchain to seed its PRNG for peer election.
Adversaries are thus tempted to influence the seed by interfering with the on-chain data to increase their chances of infiltrating anonymity services.
Our rationale for \name's robustness only holds if we can effectively prevent such interference.
As we described in Section~\ref{sec:design:spawning}, we include user-submitted entropy into the seed derivation to ensure that seeds are not entirely determined by the miners of \name's host blockchain.
However, by incorporating the spawn block, we, in return, drastically limit the capabilities of an adversary.
Namely, the adversary must (a)~successfully mine the spawn block $S_i$ for pulse $P_i$, while (b)~crafting this block to yield, in conjunction with the user-supplied entropy, a biased pre-image of a favorable seed, which is (c)~derived from a cryptographic hash function.
Assuming that no adversary possesses the computing power to control the host blockchain, we deem this kind of attack economically infeasible as honest mining is more profitable for the adversary.
In the future, we could also adapt \name to consider multiple consecutive spawn blocks to further thwart the influence of adversaries.

\vspace{-0.5em}
\subsubsection*{\bfseries Security of Handover Process}
For most anonymity services, \name requires indirection through the connector when first establishing connections.
During this handover process, each participant's connector has to authenticate all privacy peers based on the public key previously announced in the respective peer advertisements.
Hence, users only connect to privacy peers controlled by operators that created valid and distinct peer advertisement.
The adversary thus cannot launch Sybil attacks through this indirection.

\vspace{-0.5em}
\subsubsection*{\bfseries Denial of Service (DoS)}
Due to our secure local peer selection, robust peer election, and secure handover primitives, the security of utilizing bootstrapped anonymity services only depends on the security guarantees offered by those services.
While \name prevents adversaries from infiltrating anonymity services with high probability, distributed services are still prone to DoS attacks, effectively preventing proper anonymization.
However, we argue that the anonymity services currently covered by \name can cope with such attacks:
First, \name allows for the efficient creation of circuits for anonymity networks.
Hence, the limited influence of single stalling relays does not significantly impede the users' privacy.
Second, CoinParty, our investigated cryptotumbler, detects and excludes stalling peers as long as adversaries did not infiltrate at least $1\slash3$ of the peers of the CoinParty instance's mixing network~\cite{2018_ziegeldorf_coinpartyv2}.
Finally, while traditional shuffling networks do not provide protection against DoS attacks, extending them with the measures taken by CoinParty achieves the same level of protection.
Thus, our peer election does not directly thwart DoS attacks, but their impact on our considered anonymity services is highly limited.

\vspace{-.7em}
\subsubsection*{\bfseries Takeaway}
In conclusion, the peer election yields trustworthy anonymity services as long as the majority of eligible privacy peers contribute honestly to providing these services, which we ensure through our Sybil-resistant peer repository and operator incentives.

\vspace{-0.7em}
\section{Performance Evaluation}%
\label{sec:eval}

We demonstrate \name's feasibility by discussing its required synchronization times and its impact on its host blockchain.

\vspace{-0.8em}
\subsection{Time Overheads}
\label{sec:eval:time}

To continually monitor \name's state, participants should maintain a local copy of its host blockchain.
However, we only rely on the correctness of the host blockchain's PoW as \name's trust anchor.
Hence, while constrained devices may rely on a trusted source to provide a correct state, \eg a trusted IoT gateway~\cite{2014_henze_gateway}, more powerful devices preferably maintain their \name state themselves.
We again consider Bitcoin as our working example for a host blockchain and highlight how initial synchronization with \name differs from a full synchronization with Bitcoin.
Since the validity of peer advertisements typically expires in \name, as with block-pruning approaches~\cite{2020_matzutt_coinprune} participants only have to download and verify the chain of Bitcoin's block \emph{headers} and process only a few recent full blocks to derive their state by searching for \name-related \code{OP\_RETURN} transactions.
The required number of blocks to process depends on the pulse length $L_p$, the validity period of single peer advertisements, and the maximum lifespan of bootstrapped anonymity networks.
Even if bootstrapped services remain active indefinitely (\cf Section~\ref{sec:use-cases:outsource}), new users can still start synchronizing from only recent blocks and afterward discover older services from the remaining blocks in the background.
After this initial synchronization, participants actively monitor the host blockchain for new \name messages.
This overhead is negligible for Bitcoin as new blocks are only mined every ten minutes~\cite{2016_tschorsch_bitcoin}.

This potentially slow block creation interval, however, introduces unavoidable delays for the bootstrapping of new anonymity services as services are only created once a pulse's spawn block has been mined (\cf Figure~\ref{fig:design:pulses}).
For instance, a pulse length of $L_p = 12$ blocks and a negotiation phase of $L_N = 3$ blocks on top of Bitcoin means that privacy peers have at most \SI{30}{\minute} to solve their PoW puzzle, but in the worst case users have to wait up to \SI{2}{\hour} until their requested anonymity service starts bootstrapping.
Regardless, our relevant use cases of shuffling networks and cryptotumblers are latency-tolerant and sometimes even deliberately stretch their operation over time to further increase the level of achieved privacy~\cite{2018_ziegeldorf_coinpartyv2}.
If more timely service utilization is required, users can consider services valid for longer periods, thereby reducing the impact of the inflicted one-time overhead.
In this case, users have to trade off delays against security as longer validity periods devalue the protection offered by periodic PoW puzzles.

Contrarily, local peer selection only depends on individual user decisions and thus only relies on knowing a recent valid state of the peer repository, \ie full synchronization up to the most recent pulse is desirable but not required.
Namely, users can instantly sample privacy peers based on their current state, and thus \name preserves the low-latency requirement of anonymity networks.

\vspace{-.7em}
\subsubsection*{\bfseries Takeaway}
We conclude that \name (a)~has low synchronization overhead, (b)~introduces feasible latencies for bootstrapping anonymity services, and (c)~supports instant local peer selection.

\begin{figure}[t]
    \centering
    \includegraphics{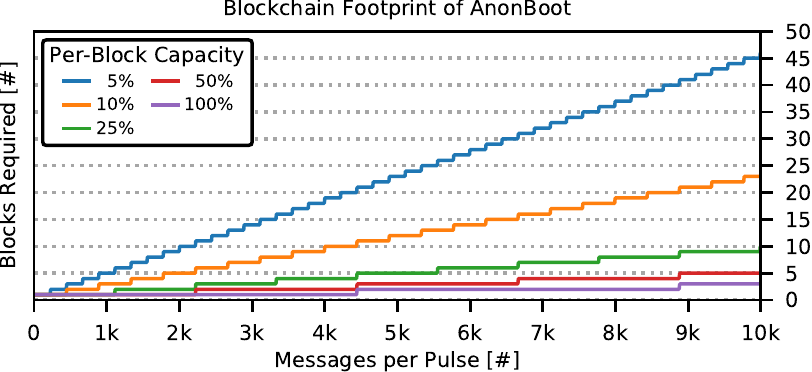}
    \Description[For increasing per-block capacities, we investigate the number of Bitcoin blocks required to publish increasing numbers of messages within one pulse.
    The required number of blocks grow linearly with the number of messages per pulse.
    In the worst case assumed for our experiment, \ie a low per-block capacity of only 5\% and 10000 messages to process per pulse, the negotiation phase can conclude after mining 45 blocks.]{
    This figure shows our evaluation of \name's blockchain footprint, using our Bitcoin-based proof-of-concept prototype.
    To assess the blockchain footprint, we assume that increasing numbers of \name messages need to be processed each pulse.
    In total, we consider up to 10000 messages per pulse and consider per-block capacities of 5\%, 10\%, 25\%, 50\%, and 100\%.
    The required number of blocks to process all messages grows linearly with the number of considered messages, and when reserving 25\% or more of each Bitcoin block for \name messages, we can process even 10000 messages within ten blocks or fewer.
    When reserving 10\% of the block size for \name messages, the negotiation phase can be concluded in under 25 blocks, and considering a per-block capacity of only 5\%, \name needs 45 blocks to process all 10000 messages.
    }
    \caption{
        \name scales to thousands of messages per pulse with only small impact on Bitcoin as its exemplary host blockchain even for constrained per-block capacities.
    }
    \label{fig:performance:blocks}
    \vspace{-0.8em}
\end{figure}

\vspace{-.8em}
\subsection{Small Blockchain Footprint and Low Costs}
\label{sec:eval:footprint}

We now show that \name realizes lightweight service discovery~\gref{goal:discovery} by assessing its impact on the host blockchain.
We measure the blockchain footprint of \name using Bitcoin's regression test mode.
As we discussed in Section~\ref{sec:design:advertisements}, the per-block capacity~$c$ helps to trade off how much transaction bandwidth \name consumes and the duration~$L_N$ of the negotiation phase.

In Figure~\ref{fig:performance:blocks}, we illustrate how the minimal required $L_N$ grows depending on the number of used \name messages and the capacity $c$.
On average, an \code{OP\_RETURN} transaction for one peer advertisement or user request has a size of \SI{307}{\byte} and a weight of \SI{901}{\btcweight} (weight units).
Since the introduction of Segregated Witnesses, the notation of block weight (limit \SI{4}{\million\btcweight}) superseded the old measure of the block size (limit \SI{1}{\mega\byte})~\cite{2017_bitcoin_segwit}.
For our measurements, we fill Bitcoin blocks while respecting the allowed capacity $c$.
Our results reveal that \name easily scales to large peer repositories and user bases with only a small footprint on Bitcoin.
When using a per-block capacity of only \SI{10}{\percent}, \name can support up to \num{10000} messages during a negotiation phase with $L_N=23$.

Peer repositories of size \num{1000}, which are already sufficiently secure as we demonstrated in Section~\ref{sec:security:selection}, have an only negligible impact on Bitcoin:
For a small capacity of only \SI{5}{\percent} to account for Bitcoin's low overall transaction throughput the negotiation phase still concludes after $L_N=5$ blocks with space for up to \num{109} user requests.
We expect only a few user requests as a single request suffices to bootstrap a service.
The scalability then only depends on the upper-layer protocol used and is independent of \name.

Finally, we briefly consider the costs inflicted by fees when privacy peers and users publish their \code{OP\_RETURN} transactions to leverage Bitcoin's consensus properties.
Albeit fluctuating, the current (March~8, 2020) recommended fee is \SI{6}{\satoshi} per byte (\SI{1}{\satoshi} $=\!10^{-8}$ \si{\bitcoin})~\cite{earn_btcfees}, and Bitcoin's market price is around \SI{9067}{\usd}~\cite{blockchain_com_charts_marketprice}.
Hence, a peer advertisement currently costs an operator \SI{0.17}{\usd}, which \name can amortize through larger pulse lengths while keeping the negotiation phase, \eg of multiple days.

\vspace{-.5em}
\subsubsection*{\bfseries Takeaway}
Overall, our analysis shows that \name can bootstrap over \num{100} services from a peer repository of size \num{1000}, serving potentially thousands of users, and can scale well beyond this size with only a small impact on Bitcoin as its host blockchain.

\vspace{-.8em}
\section{Related Work}
\label{sec:related-work}

The bootstrapping problem and Sybil attacks are inherent for distributed protocols.
In 2007, Knoll \etal~\cite{2007_knoll_bootstrapping} surveyed different approaches to finding entry points for established peer-to-peer networks.
Among other approaches, the authors proposed to bootstrap nodes through a distributed host system such as IRC~\cite{2007_knoll_bootstrapping}.
Orthogonally, Levine \etal~\cite{2006_levine_sybil} reviewed approaches to mitigate Sybil attacks.
From this taxonomy, only resource testing and recurring costs and fees are applicable to fully decentralized systems without further assumptions.
Recurring costs, namely periodic PoW-based refreshments of eligibility, are a familiar building block in the field of blockchain sharding~\cite{2016_luu_elastico,2018_zamani_rapidchain,2018_kokoris_omniledger}, where responsibilities to verify the proposed transactions are distributed among full nodes over time to improve scalability.
\name adapts this Sybil-resistant building block in the form of peer advertisements to implement the novel application that is securely bootstrapping distributed anonymity services.
In line with Knoll \etal~\cite{2007_knoll_bootstrapping}, we publish these periodic advertisements through a public blockchain as \name's host system and trust anchor.
This choice allows us to massively reduce the coordination complexity in \name since blockchains already offer a distributed means to reach a consensus of state.

Recently, Lee \etal~\cite{2018_lee_bootstrapping_isp} proposed that the user's ISP could provide privacy services, such as address hiding or VPN tunneling.
This work is orthogonal to our approach as we bootstrap services without relying on a dedicated central operator.
Namely, \name can also help users to increase their privacy against the ISP itself.

As one of its applications, \name realizes a decentralized directory service for anonymity networks such as Tor.
Similar contributions were made by other works, \eg NISAN~\cite{2009_panchenko_nisan} or ShadowWalker~\cite{2009_mittal_shadowwalker}.
However, while both proposals prevent adversarial bias, they do not feature \name's protection against Sybil attacks.
Furthermore, these approaches do not address the challenges of heterogeneous privacy peers, such as Tor nodes with different exit policies.
\name introduces the capabilities in its peer advertisements specifically to overcome this shortcoming.
While approaches to realize sticky data policies on how to handle user privacy~\cite{2011_pearson_sticky} are related to our specification of peer capabilities, even highly compressed policies such as provided by CPPL~\cite{2016_henze_cppl} may exceed our space limitations, especially when relying on Bitcoin's \code{OP\_RETURN} transactions to operate \name.
Although CPPL may facilitate simple peer capabilities, more complex instances, such as Tor relay descriptors, require manual capability abstractions.

\vspace{-0.6em}
\section{Conclusion}
\label{sec:conclusion}

We introduced \name, a blockchain-based medium to securely bootstrap distributed anonymity services via already established public blockchains, such as Bitcoin, as a trust anchor.
All \name peers communicate with each other through on-chain transactions, and, thereby, they are able to derive the same local view on \name's state by simply monitoring the host blockchain.
Our design allows for discovering peers to create Tor circuits as well as to bootstrap shuffling networks and distributed cryptocurrency tumblers on demand.
\name achieves its resilience against adversaries by two core mechanics:
First, Sybil attacks are thwarted by forcing peers to periodically refresh their membership in a repository of peers who are eligible to provide anonymity services while including a memory-bound, and thus fair, proof of work.
Second, an adversary who joins this peer repository cannot bias the peer election for new anonymity services since this peer election is based on user inputs as well as future blocks from the host blockchain.

The evaluation of our Bitcoin-based prototypic implementation of \name shows that public blockchains constitute a well-suited foundation for bootstrapping distributed systems:
\name can easily maintain a peer repository consisting of \num{1000} peers on top of Bitcoin, managing services for potentially thousands of users.
These results show that \name can operate on top of most blockchains, even if they have limited capabilities to store application-level data.

In the future, \name's utility can be further increased by identifying novel use cases apart from anonymity services.
\name lends itself to bootstrapping any distributed service, \eg to distribute trust in other domains via secure multiparty computation.

\vspace{-0.6em}
\begin{acks}
This work has been funded by the German Federal Ministry of Education and Research (BMBF) under funding reference numbers 16DHLQ013 and Z31 BMBF Digital Campus.
The funding under reference number Z31 BMBF Digital Campus has been provided by the German Academic Exchange Service (DAAD).
The responsibility for the content of this publication lies with the authors.
The authors thank J{\"o}ran Wiechert for his support with the prototype.
\end{acks}

\vspace{-0.6em}
\bibliographystyle{ACM-Reference-Format-limit}
\balance
\bibliography{paper}

\end{document}